\documentclass[runningheads]{llncs}

\usepackage[T1]{fontenc}
\usepackage{graphicx}
\usepackage{amsmath,amssymb}
\usepackage{booktabs}
\usepackage{multirow}
\usepackage{xcolor}
\usepackage{hyperref}
\usepackage{cleveref}
\usepackage{listings}
\usepackage{subcaption}
\usepackage{booktabs}
\usepackage{algorithm}
\usepackage{algpseudocode}
\usepackage{listings} 
\usepackage{listingsutf8}
\usepackage{tikz}
\usepackage{longtable}
\usetikzlibrary{arrows.meta, positioning, shapes.geometric, fit}

\lstdefinelanguage{Lean4}{
  morekeywords={theorem, lemma, def, import, namespace, end, by, sorry,
                induction, simp, exact, rfl, omega, obtain, rcases,
                intro, apply, rw, have, show, with, fun, let, where,
                variable, instance, class, structure, abbrev, open,
                if, then, else, match, do, return, section},
  sensitive=true,
  morecomment=[l]{--},
  morecomment=[s]{/-}{-/},
  morestring=[b]",
  literate={→}{{$\rightarrow$}}1 {←}{{$\leftarrow$}}1
           {↔}{{$\leftrightarrow$}}1 {∀}{{$\forall$}}1
           {∃}{{$\exists$}}1 {¬}{{$\neg$}}1
           {≤}{{$\leq$}}1 {≥}{{$\geq$}}1
           {ℕ}{{$\mathbb{N}$}}1 {α}{{$\alpha$}}1
           {β}{{$\beta$}}1 {γ}{{$\gamma$}}1
           {ε}{{$\varepsilon$}}1 {⬝}{{$\cdot$}}1
           {∧}{{$\land$}}1 {⟨}{{$\langle$}}1 {⟩}{{$\rangle$}}1
           {ₚ}{{$_p$}}1 {ₛ}{{$_s$}}1 {·}{{$\cdot$}}1,
}

\begin{document}

\title{LAMP: Lean-based Agentic framework with MCP and Proof Repair}
\author{Santhana Srinivasan R\inst{1} \and Maithilee Patawar\inst{1}}
 \institute{Indian Institute of Information Technology, Design and Manufacturing, Kancheepuram }

\maketitle

\begin{abstract}
Large language models are increasingly capable of mathematical reasoning, but the proofs they generate are often unreliable and hard to verify. Interactive theorem provers such as Lean~4 address this by accepting only kernel-checked proofs; however, their reach is bounded by the formalized knowledge available. While Mathlib, a repository of formalized Lean 4 theorems that covers diverse mathematical areas, certain specialized areas remain underrepresented; notably, the domain of Combinatorics on Words (CoW). CoW studies sequences, exploring their properties such as periodicity, borders, conjugacy, and morphisms. As a result, specialized provers, trained on Mathlib-centered data, lack the lemmas to operate in CoW. We present two contributions. First, we introduce a Lean~4 formalization of CoW containing eight modules and \textbf{93} declarations of core definitions and foundational lemmas. Second, we present LAMP, a multi-agent framework that synthesizes kernel-verified Lean~4 proofs by providing explicit, structured domain knowledge at inference time through an ontology, rather than by fine-tuning a prover. LAMP coordinates a Planner, Builder, and Verifier with Model Context Protocol based access to a domain-specific CoW ontology. In a suite of \textbf{90} CoW theorems that span all eight modules and three difficulty levels, LAMP synthesizes verified proofs for \textbf{96.7\%} of theorems, substantially exceeding both an unscaffolded baseline and existing specialized provers. An ablation shows that removing LAMP's tool-grounded architecture or its Planner/Builder separation each cost roughly 12 percentage points, even with the backbone model held fixed. 

\keywords{Theorem Proving \and Lean 4 \and LLM Agents \and
Combinatorics on Words \and 
Model Context Protocol \and Proof Automation }
\end{abstract}

\section{Introduction}
\label{sec:intro}
 
Large language models (LLMs) have made rapid progress in mathematical reasoning, but the solutions they produce are often difficult to verify and unreliable \cite{deepseekproverv2,goedelproverv2,malot}. Interactive theorem provers such as Lean~4 \cite{lean4} offer a reliable solution to the problem. With Lean~4, every proof is checked by a small trusted kernel, and the kernel-verified proof is correct regardless of how it was generated. This makes the pairing of LLM-based reasoning with formal verification a promising route to scaling the formalization of mathematics, which is a task that has traditionally demanded enormous expert effort \cite{leroy2016compcert}.

The usefulness of this approach depends on the availability of formalized domain knowledge. The Lean~4 ecosystem is built on Mathlib \cite{mathlib}, a large library of formalized mathematics. It covers domains like analysis, number theory, linear algebra, and formal verification. However, its coverage is far from complete, and several domains are unrepresented. One such domain is \emph{Combinatorics on Words} (CoW) \cite{lothaire}, the study of structural properties of finite and infinite sequences. These properties include periodicity, borders, conjugacy, morphisms, and primitive roots, which have no formalization in Mathlib. This gap is important because proofs in CoW are often long and involve many cases. Such proofs benefit from machine checking, but without a formal library, both human formalizers and automated provers lack the basic definitions needed to build these proofs.
 
Existing automated provers do not close this gap. State-of-the-art (SOTA) systems, whether fine-tuned for Lean (DeepSeek-Prover-V2 \cite{deepseekproverv2}, Goedel-Prover \cite{goedelproverv2},
Kimina-Prover \cite{kiminaprover}) are trained on a fixed, Mathlib-centered distribution. Their performance is known to decline on tasks formalized in unfamiliar frameworks \cite{taobench}.
For a domain absent from Mathlib, such as CoW, this degradation is acute. These systems lack both the vocabulary and the supporting lemmas needed to reason about CoW concepts. Therefore, the main limitation is not the inherent difficulty of the reasoning task, but the ability of these systems to generalize to new domains.

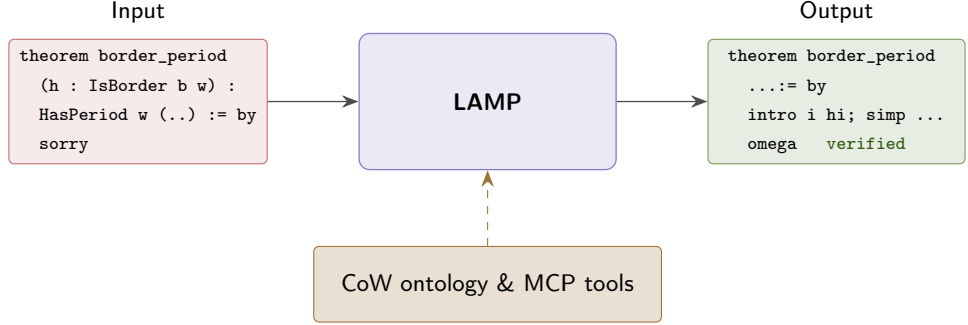
\begin{figure}[t]
\centering
\begin{tikzpicture}[
  font=\sffamily\small,
  >={Stealth[length=2.2mm]},
  node distance = 6mm,
  every node/.style = {align=center, font=\sffamily\small},
  io/.style       = {rectangle, rounded corners=2pt, draw, line width=0.4pt,
                     inner sep=4pt, minimum width=34mm, minimum height=14mm,
                     font=\sffamily\footnotesize\ttfamily, align=left},
  lamp/.style     = {rectangle, rounded corners=4pt, draw, line width=0.5pt,
                     fill=lamp@purple!12, draw=lamp@purple!70,
                     minimum width=34mm, minimum height=18mm,
                     font=\sffamily\bfseries},
  tool/.style     = {rectangle, rounded corners=2pt, draw, line width=0.4pt,
                     fill=lamp@amber!18, draw=lamp@amber!80,
                     minimum width=46mm, minimum height=10mm,
                     font=\sffamily\footnotesize},
  inputbox/.style = {io, fill=lamp@red!10, draw=lamp@red!70},
  outputbox/.style= {io, fill=lamp@green!12, draw=lamp@green!70},
  arr/.style      = {->, line width=0.5pt, draw=black!70},
  tarr/.style     = {->, line width=0.4pt, draw=lamp@amber!80, dashed}
]
 
\definecolor{lamp@red}{HTML}{A32D2D}
\definecolor{lamp@purple}{HTML}{534AB7}
\definecolor{lamp@green}{HTML}{3B6D11}
\definecolor{lamp@amber}{HTML}{854F0B}
 
\node[inputbox] (in) {%
  \scriptsize theorem border\_period\\
  \scriptsize \ \ (h : IsBorder b w) :\\
  \scriptsize \ \ HasPeriod w (..) := by\\
  \scriptsize \ \ sorry
};
 
\node[lamp, right=12mm of in] (lamp) {LAMP};
 
\node[outputbox, right=12mm of lamp] (out) {%
  \scriptsize theorem border\_period\\
  \scriptsize \ \ \dots := by\\
  \scriptsize \ \ intro i hi; simp \dots\\
  \scriptsize \ \ omega \quad {\color{lamp@green!80!black}\textbf{verified}}
};
 
\node[tool, below=10mm of lamp] (tools) {CoW ontology \& MCP tools};

\node[font=\sffamily\footnotesize, above=1mm of in]   {Input};
\node[font=\sffamily\footnotesize, above=1mm of out]  {Output};

\draw[arr] (in)    -- (lamp);
\draw[arr] (lamp)  -- (out);
\draw[tarr] (tools) -- (lamp);
 
\end{tikzpicture}
\caption{LAMP transforms a theorem with \texttt{sorry} into a kernel-verified proof}
\label{fig:overview}
\end{figure}

Our main observation is that an unformalized domain needs explicit and structured domain knowledge at inference time, rather than knowledge that is only learned implicitly through training. This knowledge includes the definitions of the domain and the relationships between them. It can be provided to a general-purpose model through external tools. In this way, the model can be adapted to a new domain by organizing and providing the required knowledge instead of retraining the model. We apply this idea specifically to CoW. Figure~\ref{fig:overview} illustrates this concretely: a CoW theorem with a \texttt{sorry} placeholder is transformed into a kernel-verified proof by a general-purpose model drawing on a curated CoW knowledge source.

We make this idea concrete through two artifacts: a Lean~4 formalization of CoW and \emph{LAMP}, a multi-agent framework for generating verified Lean~4 proofs. LAMP uses curated domain knowledge during the proving process via Model Context Protocol (MCP) \cite{mcp}. It combines three main agents: a Planner, which creates a proof strategy; a Builder, which turns the strategy into Lean tactics; and a Verifier, which checks the proof against the Lean kernel. Most importantly, these agents can access a domain-specific CoW ontology that provides the required definitions and their dependencies when needed. We summarize our contributions as
follows:
 
\begin{enumerate}
  \item The first Lean~4 formalization of CoW. It contains eight modules (Word, Factor, ProperPrefix, ProperSuffix, Border, Conjugacy, Period, Morphism) with core definitions and foundational lemmas.
  \item LAMP, a multi-agent framework that injects domain knowledge into LLM  proof synthesis via MCP tools. It includes a structured CoW ontology that is used by a dedicated Planner agent. This allows the framework to adapt to the CoW domain without retraining the underlying model.
  
  \item A CoW Evaluation Suite of \textbf{90} theorems across the eight modules
and three difficulty levels, including the classical border-period duality
theorem, designed as a growing foundation: solved theorems can be folded back
into the library to support harder results, rather than serving as a static
benchmark.
\end{enumerate}
 
On this suite, LAMP synthesizes verified proofs for \textbf{96.7}\% of theorems,
substantially exceeding both an unscaffolded LLM baseline (\textbf{58.9}\%) and
existing specialized provers (\textbf{1.1}--\textbf{8.9}\%), which fail almost
entirely because CoW lies outside their training distribution. The ablation
above shows that this performance depends on LAMP's tool-grounded architecture
and its Planner/Builder separation, not on the backbone model alone. We also
report performance on an out-of-domain benchmark, where LAMP is deliberately
less competitive: it trades breadth across general mathematics for depth in an
unformalized domain, and we make this trade-off explicit rather than claiming
general superiority.
 
The rest of the paper is organized as follows. Section~\ref{sec:related}
reviews related work in theorem proving and agentic systems. Section~\ref{sec:lamp}
introduces the Lean~4 formalization of CoW (Section~\ref{sec:cow}) and then
describes the LAMP system, its components, and control algorithm.
Section~\ref{sec:eval} reports the evaluation, ablations, and comparisons.
Section~\ref{sec:discussion} opens with a failure analysis and then discusses
implications, limitations, and future work, and Section~\ref{sec:conclusion}
concludes.

\section{Background and Related Work}
\label{sec:related}

Three lines of work are directly related to our setting. We first review the proof assistant our system targets and the library that anchors it (Section \ref{sec:related-lean}). We then survey approaches that leverage large language models for Lean proof generation (Section \ref{sec:related-provers}), and finally agentic frameworks that augment general-purpose models with external tools to the same end (Section \ref{sec:related-agentic}). A comprehensive recent survey by \cite{dlsurvey} provides broader context; here we focus on the aspects most salient to the design and evaluation of LAMP.

\subsection{Lean~4, Mathlib, and Combinatorics on Words}
\label{sec:related-lean}

Lean~4 is an interactive theorem prover (ITP) and dependently typed functional
programming language in which mathematical statements are encoded as types and
proofs as terms whose well-typedness is checked by a small trusted
kernel~\cite{lean4}. Proofs are usually developed interactively through
\emph{tactics}, which transform a proof state consisting of hypotheses and goals
until the goal is discharged. This design makes Lean~4 an attractive target for
machine-generated proofs: any candidate proof, however it is produced, is
accepted only if it passes kernel checking, so correctness does not depend on
trusting the generator. Other widely used proof assistants include
Coq~\cite{coq} and Isabelle/HOL~\cite{isabelle}, though the overwhelming
majority of recent LLM-based prover research targets Lean~4.

Lean's utility relies heavily on Mathlib~\cite{mathlib}, a community-driven
library providing an extensive network of definitions, lemmas, and automation
across diverse mathematical fields, including algebra, analysis, topology,
number theory, and combinatorics. Mathlib is also a fast-moving target:
definitions are routinely added, renamed, or deprecated, and proof developments
must track these changes. Despite its breadth, Mathlib's coverage is uneven,
and several mathematical domains remain entirely unformalized.

\emph{Combinatorics on Words}
(CoW)~\cite{lothaire,lothaire2002algebraic} is one such domain. CoW studies
the structural and combinatorial properties of finite and infinite sequences,
encompassing periodicity, borders, conjugacy, primitive roots, and morphisms.
CoW hosts several long-standing open problems that have driven sustained
research activity. A central question, posed by Fraenkel and
Simpson~\cite{fraenkelsimpson1998}, asks how many distinct squares a word of
length~$n$ can contain; the conjectured upper bound of~$n$ remains open, with
the best known bound of fewer than~$1.5n$ distinct squares established by
Thierry~\cite{thierry2020}. The closely related \emph{runs conjecture}, which
posits that the number of maximal repetitions in a word of length~$n$ is at
most~$n$, was formulated by Crochemore et al.~\cite{runsconjecture} and
subsequently resolved. Recent advances continue to sharpen density bounds for
specific repetition patterns~\cite{patawarsquaredensity}. A recurring feature
of proofs in this area is that they involve elaborate case analyses over word
combinatorics, making them both lengthy and error-prone to verify by hand;
this characteristic creates a natural demand for machine-checked formalization.

The most substantial formalization of CoW to date is the project of Holub and
Starosta in Isabelle/HOL, which formalizes the Periodicity Lemma, the
Lyndon--Sch\"{u}tzenberger equation, the Graph Lemma, and additional results
on Lyndon words and morphic
languages~\cite{holubstarosta2021,holubstarosta2021lyndon,starosta2023morphic}.
No comparable Lean~4 formalization exists, which means that both researchers
and automated provers must build the necessary definitions and supporting
lemmas from scratch. The proposed work addresses this gap by establishing a CoW
library in Lean~4.

\subsection{LLM-based Theorem Provers}
\label{sec:related-provers}

The application of neural language models to theorem proving originates with GPT-f~\cite{gptf}, which demonstrated that a transformer fine-tuned on proof data can generate single proof steps in Metamath, guided by best-first search. Subsequent work scaled this paradigm: PACT~\cite{pact} introduced proof artifact co-training, while curriculum learning over statement difficulty further improved success rates on formal mathematics~\cite{curriculumlearning}. HyperTree Proof Search (HTPS)~\cite{htps} brought AlphaZero-style online training to theorem proving, achieving strong results on both Metamath and the Lean-based miniF2F benchmark.

More recently, several systems have fine-tuned large language models specifically for whole-proof or stepwise proof generation in Lean. The DeepSeek-Prover series~\cite{deepseekprover,deepseekproverv15,deepseekproverv2} couples whole-proof generation with subgoal decomposition and reinforcement learning, with its largest model reaching SOTA results on competition benchmarks. Kimina-Prover~\cite{kiminaprover} adopts a reasoning-driven reinforcement-learning framework. In contrast, the Goedel-Prover line~\cite{goedelprover,goedelproverv2} achieves comparable performance with substantially smaller models by pairing scaffolded data synthesis with verifier-guided self-correction. InternLM2.5-StepProver~\cite{internlmstepprover} scales expert iteration on the Lean Workbook dataset~\cite{leanworkbook} using a learned critic model to guide search toward tractable subproblems. LEGO-Prover~\cite{legoprover} constructs reusable lemma libraries during proof search, and Baldur~\cite{baldur} demonstrates whole-proof generation with repair in Isabelle. Draft, Sketch, and Prove~\cite{dsp} bridges informal and formal reasoning by using an LLM to produce natural-language proof sketches that are then translated into formal tactic sequences. AlphaProof~\cite{alphaproof}, combining reinforcement learning with autoformalization and Lean~4 verification, achieved silver-medal performance at the 2024 International Mathematical Olympiad, marking a landmark for AI-driven formal mathematics.

These systems are typically evaluated on miniF2F~\cite{minif2f} and PutnamBench~\cite{putnambench}, both of which are derived from mathematics competitions. ProofNet~\cite{proofnet} extends evaluation to undergraduate-level mathematics, and FormalMATH~\cite{formalmath} provides a broader benchmark across multiple mathematical domains.

A complementary family augments LLMs with retrieval. LeanDojo and its ReProver model~\cite{leandojo} extract training data from Lean repositories and perform premise selection over Mathlib, retrieving relevant lemmas to condition proof generation. Lean Copilot~\cite{leancopilot} integrates LLM-based tactic suggestion directly into the interactive Lean workflow. Autoformalization, the automatic translation from natural language mathematics to formal specifications, has also emerged as a closely related direction, with large language models showing promising capabilities for this task~\cite{autoformalization}.

Despite strong benchmark numbers, these approaches share limitations that are salient for our setting. Because they are trained predominantly on competition mathematics and on fixed snapshots of Mathlib, their ability to generalize to domains outside their training distribution is unclear, and they can be brittle to Mathlib version changes such as renamed or deprecated lemmas~\cite{axprover}. Recent benchmarks specifically probe this concern: TaoBench~\cite{taobench} asks whether automated prover LLMs generalize beyond Mathlib, miniCTX~\cite{minictx} evaluates provers on novel projects with long contexts, and related work observes that the effectiveness of specialized provers diminishes on tasks involving unfamiliar problem domains~\cite{leangeo}. Finally, the largest specialized provers require substantial computational resources to deploy. These observations motivate approaches that introduce domain knowledge without retraining a monolithic model, which is the direction our work pursues.

\subsection{Agentic Frameworks for Theorem Proving}
\label{sec:related-agentic}

An alternative paradigm equips general-purpose LLMs with external tools and coordinates them through agentic orchestration, rather than fine-tuning a dedicated prover. This builds on broader work in tool-augmented language models~\cite{react,toolformer}. In the theorem-proving setting, the LLM supplies reasoning and Lean expertise, while tools provide access to the proof state, lemma search, error diagnostics, and verification. Several systems explore agent decompositions of the proving task, including multi-agent long-chain-of-thought reasoning~\cite{malot}, agent-based proof refinement~\cite{proveragent}, interleaving of informal reasoning steps with tactic generation~\cite{leanstar}, and language-agent approaches that couple reasoning with environment interaction~\cite{langagent}. MCP~\cite{mcp} has recently emerged as a standard interface for connecting
LLMs to such tools, including Lean language-server tooling~\cite{leanlspmcp}.

LAMP shares its agentic, MCP-based philosophy with Ax-Prover \cite{axprover},
which also pairs a frontier LLM with Lean tooling and argues that tool-augmented
agents generalize where specialized provers do not; we view our work as
complementary rather than competing, and differ in three respects we make
explicit and evaluate. First, Ax-Prover demonstrates breadth across mathematics and quantum physics using generic Lean tooling. In contrast, LAMP targets CoW, a domain absent from both Mathlib and prior agentic benchmarks. To address this, LAMP contributes a dedicated CoW Lean~4 library as a reusable dataset. No generic Lean tool can supply this knowledge, as it does not yet exist outside our work. Second, whereas existing agentic provers rely on
generic Lean tooling, LAMP introduces a structured, domain-specific CoW \emph{ontology} that is exposed as an MCP tool to Builder agent and also consumed by a dedicated Planner agent during proof-strategy formation, injecting curated domain knowledge into planning rather than leaving it implicit in the model's weights. Third, LAMP separates the planning of a proof strategy from the construction of Lean tactics into distinct Planner and Builder agents, coordinated by a
dual-loop orchestrator. In Section~\ref{sec:eval}, we show that both LAMP's tool-grounded architecture and its Planner/Builder separation are necessary for its performance on this unformalized domain. Removing either component produces a marked drop in the proof-synthesis rate. This isolates a unique contribution that generic tool-augmented agents do not provide.

\section{The LAMP System}
\label{sec:lamp}
\subsection{The CoW Library}
\label{sec:cow}
The library models words as lists over an arbitrary type (\texttt{abbrev Word ($\alpha$ : Type*) := List $\alpha$}) and is organized into eight modules with a strict dependency hierarchy: \emph{Word} $\to$ \emph{Factor} $\to$ \{\emph{ProperPrefix}, \emph{ProperSuffix}\} $\to$ \emph{Border}, with \emph{Conjugacy}, \emph{Period}, and \emph{Morphism} each built directly atop \emph{Word} and \emph{Factor}; later modules import only earlier ones. \emph{Word} establishes the monoid structure under concatenation. \emph{Factor} formalizes prefix, suffix, and factor relations with full transitivity chains; \emph{ProperPrefix} and \emph{ProperSuffix} capture strict subword relations and length inequalities. \emph{Border} encodes words that simultaneously appear as a proper prefix and suffix, illustrated below: a border is defined directly in terms of the two prior modules, and its core closure lemma follows by chaining a single prefix-to-factor lemma from \emph{Factor}. The following Lean~4 snippet illustrates the \textit{Border} definition and its entailment of the factor relation:

\begin{lstlisting}
def IsBorder (b w : Word α) : Prop :=
  IsProperPrefix b w ∧ IsProperSuffix b w

theorem IsBorder.isFactor (h : IsBorder b w) : IsFactor b w :=
  IsFactor.of_prefix h.1.isPrefix
\end{lstlisting}

\emph{Conjugacy} characterizes cyclic rotations. \emph{Period} defines periodicity, minimal periods, and primitivity. \emph{Morphism} covers word morphisms together with non-erasing, coding, and uniform classifications. Table~\ref{tab:cow-library} gives the declaration breakdown per module; additional implementation details and design rationale are given in Appendix~\ref{app:cow}.

\begin{table}[t]
\centering
\caption{Declaration breakdown of the CoW Lean~4 library by module.}
\label{tab:cow-library}
\begin{tabular}{lccc}
\toprule
Module & Definitions & Lemmas & Total \\
\midrule
Word         &  4 & 18 & 22 \\
Factor       &  4 & 18 & 22 \\
ProperPrefix &  1 &  4 &  5 \\
ProperSuffix &  1 &  4 &  5 \\
Border       &  3 &  4 &  7 \\
Conjugacy    &  1 &  3 &  4 \\
Period       &  3 &  6 &  9 \\
Morphism     &  5 & 14 & 19 \\
\midrule
\textbf{Total} & \textbf{22} & \textbf{71} & \textbf{93} \\
\bottomrule
\end{tabular}
\end{table}

\subsection{System Overview}
\label{sec:lamp-overview}
 
\begin{figure}[t]
  \includegraphics[width=\linewidth]{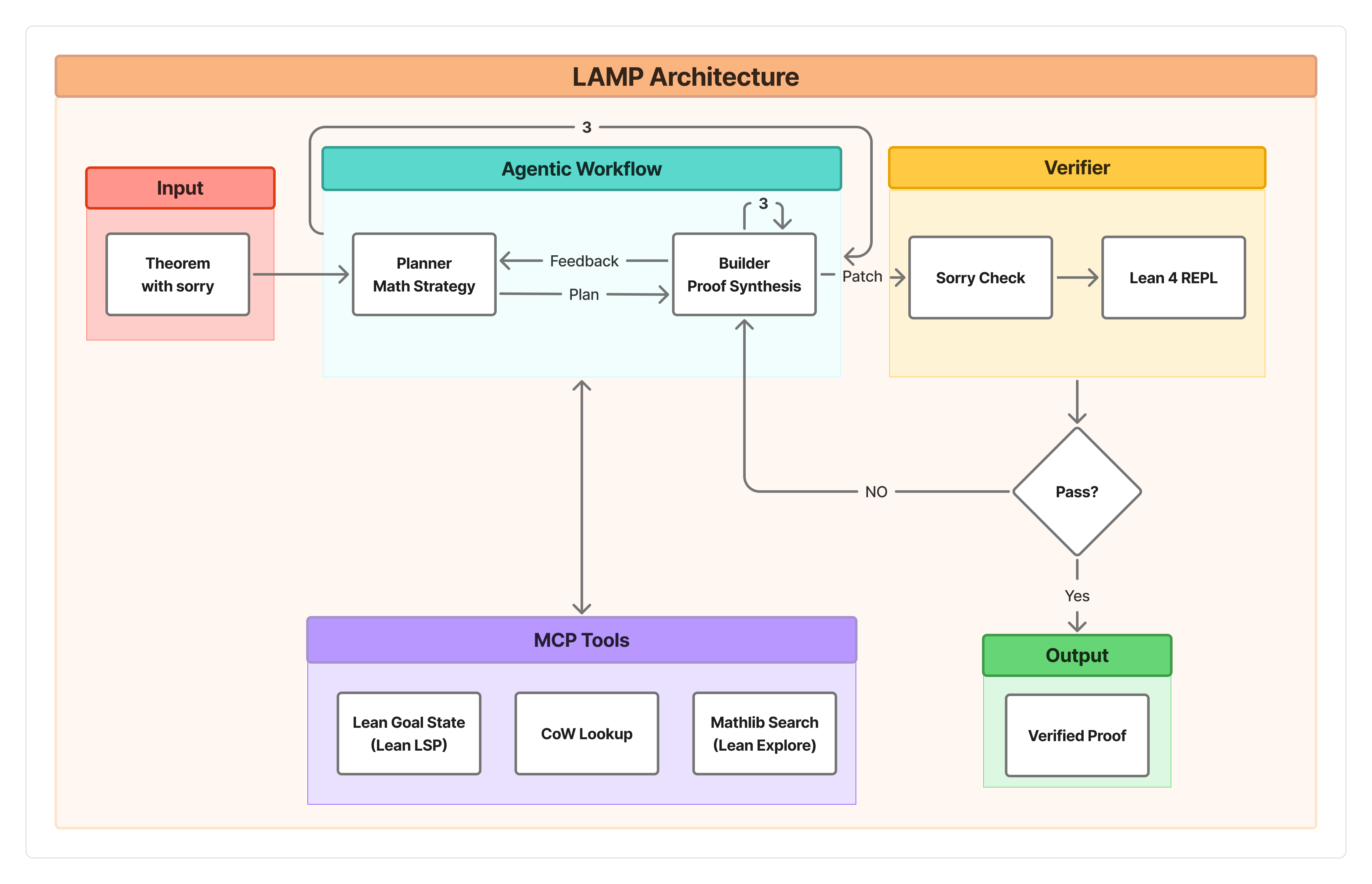}
   \caption{LAMP architecture and data flow}
 \label{fig:architecture}
 \end{figure}

LAMP is a multi-agent framework that synthesizes a complete, kernel-verified
Lean~4 proof from a theorem statement whose proof body is left as a
\texttt{sorry} placeholder. Its design follows a single principle: rather
than asking one language model to perform mathematical reasoning, Lean coding,
and error recovery simultaneously, LAMP decomposes the task across specialized
agents and grounds each of them in external tools so that decisions are based
on verified facts rather than the model's recollection. Three core agents cooperate in this architecture. A \emph{Planner} devises a mathematical proof strategy, a \emph{Builder} realizes that strategy as Lean~4 tactics, and a \emph{Verifier} checks candidate proofs against the Lean kernel. Managing this workflow is a central \emph{Orchestrator}, while a \emph{Semantic Toolbox} exposes a set of MCP \cite{mcp} tools that the agents may query on demand. These tools include a domain-specific CoW ontology. The system operates as a closed feedback loop: a strategy is planned, built into code, and verified; if verification fails, the structured error is returned to the agents, which revise and retry. This loop runs under a
bounded attempt budget until a proof is verified or the budget is exhausted. Figure~\ref{fig:architecture} shows the overall data flow. The remainder of this section describes each component (Section~\ref{sec:lamp-components}) and then specifies the control algorithm that ties them together (see Section~\ref{sec:lamp-algorithm}).
 
\subsection{Components}
\label{sec:lamp-components}
 
We describe the five components in data-flow order: the Orchestrator that
controls the loop, the Planner and Builder agents that produce strategy and
code, the Semantic Toolbox that grounds the Builder, and the Verifier that
adjudicates correctness. All agent reasoning is performed by a single
general-purpose large language model accessed through an API; LAMP is
model-agnostic, and we report results across several backbones in
Section~\ref{sec:eval}.
 
\paragraph{Orchestrator:}
The Orchestrator is the central controller that drives the
``Planner $\rightarrow$Builder$\rightarrow$Verifier" cycle and owns all loop state. Before any reasoning begins, it performs \emph{contextual grounding}: it
extracts the \texttt{import} statements from the target theorem, resolves
each to the corresponding CoW source file, and injects the verbatim library
source into both agents' prompts. This ensures the agents reason about the
library's actual definitions and lemma names rather than plausible-looking
inventions, which is the principal defense against hallucinated identifiers.
The Orchestrator also localizes edits: a theorem locator identifies the exact
source region occupied by the target proof so that generated tactics replace
only the \texttt{sorry} body and never overwrite imports or neighbouring
declarations, with a conservative fallback that patches only the lines
containing \texttt{sorry} when boundaries cannot be determined. Finally, the
Orchestrator enforces the bounded attempt budget and the branching recovery
policy formalized in Section~\ref{sec:lamp-algorithm}, and persists every
verified proof to disk.
 
\paragraph{Planner Agent:}
The Planner is the mathematical strategist and does not emit Lean code. Given
the goal and the injected library context, it produces an explicit
chain-of-thought analysis followed by a structured \emph{plan} that names the
specific library lemmas and the proof technique to apply (for example, chaining
prefix relations through a transitivity lemma before applying a factor
lemma). The Planner retains a persistent conversation history across re-planning
iterations, so that when a strategy fails it revises in light of the Builder's
accumulated feedback rather than planning from scratch. Separating strategy from
implementation in this way lets the system distinguish a flawed mathematical
approach (which warrants re-planning) from a correct approach that was
implemented incorrectly (which warrants only a rebuild).
 
\paragraph{Builder Agent:}
The Builder is the tactical executor that translates the Planner's strategy into
syntactically valid Lean~4 tactics. During generation, it queries the Semantic Toolbox (up to a fixed limit per attempt) to retrieve definitions, search for lemmas, or inspect the live goal state before synthesizing the final proof. Within a single planning iteration the Builder maintains its own history: if an attempt fails verification, the next attempt receives the structured error and the prior code, enabling targeted self-correction without re-prompting from scratch. A post-processing step strips any accidental re-emission of the theorem signature so that only the proof body is patched into the source.
 
\paragraph{Semantic Toolbox:}
The Semantic Toolbox is the tool registry that exposes the MCP
ecosystem \cite{mcp} to the Builder and routes each call to the appropriate
backend. It provides three classes of tool. A \emph{Lean LSP} tool wraps the Lean language server \cite{leanlspmcp} to extract the precise goal state at the location of a \texttt{sorry}. This state consists of the hypotheses and target exactly as the kernel sees them. Consequently, it provides the Builder with ground truth rather than an inferred proof state.A \emph{Mathlib search} tool, built on LeanExplore \cite{leanexplore}, queries the surrounding Lean ecosystem for general-purpose lemmas. Most importantly, an \emph{ontology} tool answers queries against a curated CoW knowledge corpus, returning the exact Lean
statement, description, source location, dependencies, and related concepts for a named CoW entity, together with operations that list all lemmas in a conceptual family or return a theorem's prerequisite chain. The tools are governed by an explicit priority hierarchy enforced through prompting: the Builder consults the injected source first, then the CoW ontology for domain-specific lookups, and only as a last resort the broader Mathlib search.
This ordering highlights a core design priority in LAMP. Specifically, the architecture privileges curated domain knowledge over open-ended retrieval. The ontology tool then serves as the vehicle for bringing this knowledge into the proof construction process.
 
\paragraph{Verifier:}
The Verifier is the final arbiter of correctness and the only component that
LAMP trusts to declare success. It submits a candidate proof to a Lean~4 REPL and parses the result into structured categories: errors, warnings, open goals, and \texttt{sorry} occurrences. This distinction separates a proof that merely compiles (\emph{pass}: no errors) from one that is genuinely finished (\emph{complete}: no errors, no sorries, no sorry-warnings). Parsed errors are returned to the agents as structured feedback, including the offending goal state, so that recovery is informed rather than blind. The verification backend,
adapted from DeepSeek-Prover-V1.5 \cite{deepseekproverv15}, runs REPL instances as isolated worker processes with per-process resource limits and timeouts, and a scheduler reaps stale processes, allowing many candidate proofs to be checked concurrently and safely.

\subsection{Algorithm}
\label{sec:lamp-algorithm}
 
Algorithm~\ref{alg:lamp} specifies LAMP's control flow as a dual-loop procedure.
The \emph{outer loop} (the planning loop) requests a proof strategy from the
Planner and may re-plan up to $R_{\max}$ times; the \emph{inner loop} (the build
loop) asks the Builder to realize the current strategy as Lean tactics and may
retry up to $B_{\max}$ times per strategy. This yields up to
$R_{\max}\!\times\!B_{\max}$ verification attempts in total. Two design
decisions in the loop are not evident from the architecture alone and are
described after the listing.
 
\begin{algorithm}[t]
\caption{LAMP proof synthesis}
\label{alg:lamp}
\begin{algorithmic}[1]
\Require theorem $T$ with a \textsc{sorry} placeholder; library $\mathcal{L}$
\Require re-plan budget $R_{\max}$, build budget $B_{\max}$
\Ensure a verified Lean~4 proof of $T$, or \textsc{failure}
\State $C \gets \textsc{LoadContext}(T, \mathcal{L})$
  \Comment{inject source of imported CoW modules}
\State $H_P \gets \emptyset$
  \Comment{persistent Planner history}
\For{$r \gets 1$ \textbf{to} $R_{\max}$}
  \State $\mathit{plan} \gets \textsc{Planner}(T, C, H_P)$
    \Comment{mathematical strategy; no Lean code}
  \State $H_B \gets \emptyset$
    \Comment{Builder history, reset per strategy}
  \For{$b \gets 1$ \textbf{to} $B_{\max}$}
    \State $\mathit{proof} \gets \textsc{Builder}(\mathit{plan}, C, H_B,
      \textsc{Toolbox})$
      \Comment{may issue MCP tool calls}
    \If{$\textsc{ContainsSorry}(\mathit{proof})$}
      \Comment{stage 1: textual check}
      \State $H_B \gets H_B \cup \{\text{``proof still contains sorry''}\}$
      \State \textbf{continue}
    \EndIf
    \State $\mathit{res} \gets \textsc{Verify}(\mathit{proof})$
      \Comment{stage 2: Lean REPL compilation}
    \If{$\mathit{res}.\mathit{complete}$}
      \Comment{no errors, no sorries}
      \State \textbf{return} $\mathit{proof}$
    \ElsIf{$\textsc{IsMissingDefinition}(\mathit{res})$}
      \State $C \gets \textsc{ReloadContext}(T, \mathcal{L})$
        \Comment{cheap retry: re-ground, do not re-plan}
      \State $H_B \gets H_B \cup \{\mathit{res}.\mathit{errors}\}$
    \Else
      \State $H_B \gets H_B \cup \{\mathit{res}.\mathit{errors}\}$
        \Comment{feed structured error back to Builder}
    \EndIf
  \EndFor
  \State $H_P \gets H_P \cup \{\text{summary of failed attempts for plan } r\}$
    \Comment{Planner learns before re-planning}
\EndFor
\State \textbf{return} \textsc{failure}
\end{algorithmic}
\end{algorithm}
 
\paragraph{Dual-stage verification:}
A candidate proof is accepted only if it passes two independent checks
(lines~9 and~14). The first is a purely textual test that the proof contains no \textsc{sorry} token. The second compiles the proof in the Lean REPL and requires that the result be \emph{complete}: free of errors, sorries, and sorry-warnings. The textual stage is necessary because Lean accepts \textsc{sorry} as a well-typed placeholder: a proof consisting of \textsc{sorry} alone compiles without error, so a compilation pass alone would register a false
success. Requiring both stages eliminates this failure mode.
 
\paragraph{Missing-definition fast path:}
When Builder fails, LAMP distinguishes \emph{why} it failed (line~16). If the error indicates that a referenced definition or lemma could not be found, this represents an addressing failure rather than a flaw in the strategy. In this case, the Orchestrator re-injects the library context and retries the build immediately (lines~17--18). It does this rather than discarding the strategy and consuming an outer-loop re-plan. A genuine proof failure instead appends the structured error to the Builder's history (line~20) so that the next build attempt can self-correct, and only when the inner budget is exhausted does control return to the Planner (line~22), which revises the strategy using a summary of what failed. This separation keeps inexpensive recovery (re-grounding) distinct from expensive
recovery (re-planning).

\section{Experimental Results}
\label{sec:eval}
 
We evaluate LAMP to answer five questions. (Q1)~How well does LAMP synthesize
proofs in an unformalized domain? (Q2)~Does LAMP's performance come from its tools and architecture, or from the underlying language model?
(Q3)~Does the result depend on a particular model backbone? (Q4)~How does LAMP
compare with existing specialized provers on the same domain? (Q5)~How does it
behave outside its target domain? We address these in Sections~\ref{sec:eval-main}--\ref{sec:eval-ood}, with a
failure analysis following in Section~\ref{sec:discussion}.
 
\subsection{Experimental Setup}
\label{sec:eval-setup}
 
\paragraph{Evaluation suite:}
We evaluate on the CoW Evaluation Suite, a curated collection of \textbf{90}
theorems spanning all eight modules of the CoW library
(Section~\ref{sec:cow}) and stratified into three difficulty levels: easy, medium, and hard. Difficulty was assigned according to the expected proof length and reasoning depth: easy theorems are dischargeable in a few tactic steps using one or two library lemmas; medium theorems require case analysis or induction over several steps; and hard theorems require
cross-module reasoning or the discovery of non-obvious intermediate lemmas. Note that these labels reflect the complexity of the \emph{formal proof task}, not the mathematical difficulty a human prover or a mathematician would perceive.The distribution of theorems across difficulty levels and modules is given in Table~\ref{tab:suite}; the complete list of all 90 theorems with per-backbone pass/fail results is given in Appendix~\ref{app:eval}. We view this suite as a foundation rather than an endpoint: its 90 theorems validate LAMP on a CoW library that did not previously exist, and verified proofs can, after human review, be folded back in to support harder results. LAMP's role is thus not only to solve theorems but to grow the foundation it solves them over.

\begin{table}[t]
\centering
\caption{Composition of the CoW Evaluation Suite.}
\label{tab:suite}
\begin{tabular}{lccc}
\toprule
Difficulty & \#\,Theorems & Modules covered & Example \\
\midrule
Easy    & 15 & 8 & \texttt{word\_length\_pos\_of\_ne\_nil} \\
Medium & 58 & 8 & \texttt{prefix\_antisymm} \\
Hard    & 17 & 5 & \texttt{border\_period\_duality} \\
\midrule
\textbf{Total} & \textbf{90} & 8 & --- \\
\bottomrule
\end{tabular}
\end{table}

 \vspace{-10pt}
\paragraph{Success criterion:}
A theorem is counted as solved only if the synthesized proof passes LAMP's
dual-stage check (Section~\ref{sec:lamp-algorithm}): it must contain no
\textsc{sorry} token, and it must compile to a \emph{complete} state under the
Lean~4 REPL---no errors, no sorries, and no sorry-warnings. This criterion
admits no partial credit.
 
\paragraph{Metric:}
We report pass@1: a single synthesis attempt per theorem under the standard
attempt budget. We adopt pass@1 because it reflects realistic usage, in which a
practitioner runs the system once under fixed time and cost constraints rather
than sampling a proof many times in the hope that one attempt succeeds. Note
that LAMP's internal repair loop (up to $R_{\max}\!=\!3$ re-plans and
$B_{\max}\!=\!3$ build attempts per plan) operates \emph{within} a single
pass@1 attempt and is distinct from independent pass@$k$ sampling.
 
\paragraph{Configurations:}
Unless otherwise stated, LAMP uses \textbf{Kimi K2.6} (Moonshot AI) as its backbone model (API
version \texttt{kimi-k2.6}, endpoint \texttt{api.moonshot.ai/v1}) at temperature \textbf{1}
(the only value accepted by the Kimi K2.6 API). All experiments use the same Mathlib snapshot
(\texttt{v4.17.0}, pinned to commit \texttt{5269898}).
The Lean toolchain is fixed at \texttt{leanprover/lean4:v4.17.0}. Each verification attempt is given a timeout of \textbf{30}
seconds. Experiments were run on a single machine equipped with a
\textbf{13th Gen Intel Core i7-13650HX} (14 cores, 28 threads) CPU,
\textbf{24\,GB} RAM, and an \textbf{NVIDIA GeForce RTX\,4060 Laptop GPU (8\,GB VRAM)},
running \textbf{Ubuntu 24.04.3 LTS}. The builder executes up to
\textbf{9} repair attempts (3 cycles $\times$ 3 attempts per cycle) before
declaring failure.

\subsection{Main Result: CoW Proof Synthesis}
\label{sec:eval-main}
 
Table~\ref{tab:main} reports LAMP's pass@1 rate on the CoW Evaluation Suite,
stratified by difficulty. LAMP solves \textbf{87} of \textbf{90} theorems, an
overall pass rate of \textbf{96.7}\%. Performance by difficulty is \textbf{100.0}\%
(easy), \textbf{98.3}\% (medium), and \textbf{88.2}\% (hard); the per-module
breakdown appears in Table~\ref{tab:main-module} (see Appendix~\ref{app:eval}). Notably, LAMP synthesizes a verified proof of border--period duality, a named
CoW result requiring a prefix and suffix decomposition chained through the
Border, Factor, and Period modules (full proof in
Appendix~\ref{app:results}). These results indicate that LAMP is effective not only on
structural lemmas but also on theorems requiring multi-step, cross-module
reasoning.
 
\begin{table}[t]
\centering
\caption{LAMP pass@1 on the CoW Evaluation Suite, by difficulty.}
\label{tab:main}
\begin{tabular}{lccc}
\toprule
Difficulty & \#\,Theorems & Solved & Pass@1 (\%) \\
\midrule
Easy & 15 & 15 & 100.0\% \\
Medium & 58 & 57 & 98.3\% \\
Hard   & 17 & 15 & 88.2\% \\
\midrule
\textbf{Overall} & \textbf{90} & \textbf{87} & \textbf{96.7\%} \\
\bottomrule
\end{tabular}
\end{table}

\subsection{Ablation: The Role of Architecture}
\label{sec:eval-ablation}
The central claim of LAMP is that its performance is driven by its architectural design. It integrates tool-grounded access to the domain ontology via MCP and separates strategy from implementation, rather than by the backbone model alone.
We test this with two ablations, holding the $3\times3$ attempt budget and injected library source fixed in both. \emph{No MCP} removes all tool access.
\emph{Single agent} removes the Planner: the Builder alone makes all nine
calls, retaining full MCP access, but must plan and prove.

\begin{table}[t]
\centering
\caption{Ablation study. All configurations use the same backbone model and
the same 9-call attempt budget; only the indicated component is removed.}
\label{tab:ablation}
\begin{tabular}{lccccc}
\toprule
Configuration & Easy & Medium & Hard & Overall & $\Delta$ \\
\midrule
LAMP (full)              & 100.0\% & 98.3\% & 88.2\% & \textbf{96.7\%} & - \\
\;-- no MCP               & 100.0\% & 91.4\%  & 47.1\% & 84.4\% & 12.3 \\
\;-- single agent (9 Builder calls) & 100.0\% & 89.7\%  & 58.8\% & 85.6\% & 11.1 \\
\bottomrule
\end{tabular}
\end{table}

Both ablations cost a comparable 11--12 points overall, with the loss
concentrated almost entirely in Hard theorems (88.2\% down to 47.1\% and
58.8\%); Easy theorems are unaffected. This indicates that tool-grounded access
to the domain ontology and the Planner/Builder separation each contribute
substantially, and neither alone explains LAMP's performance.
  
\subsection{Backbone Comparison and Comparison with Existing Provers}
\label{sec:eval-backbone}
 
LAMP is model-agnostic: any instruction-following LLM with tool-use support can serve as the backbone for its agents. Table~\ref{tab:backbone} reports LAMP's
pass@1 on the CoW suite with the backbone varied, architecture and tools held fixed. LAMP attains 96.7\% with its backbone model, 88.9\% with Claude Sonnet 4.5, and 68.9\% with DeepSeek V4 Pro. We emphasize that this experiment varies only the \emph{engine} inside LAMP; it is distinct from the system-level comparison, where complete proving systems
compete against one another.
\begin{table}[t]
\centering
\caption{Effect of the backbone model on LAMP, CoW suite (architecture fixed).}
\label{tab:backbone}
\begin{tabular}{lc}
\toprule
Model & Pass@1 (\%) \\
\midrule
Backbone model      & \textbf{96.7\%} \\
Claude Sonnet 4.5    & 88.9\% \\
DeepSeek V4 Pro      & 68.9\% \\
\bottomrule
\end{tabular}
\end{table}
 
We now compare LAMP against specialized theorem-proving systems and an unscaffolded baseline on the same CoW suite, with the same ontology context given to every system, including the provers.

\begin{table}[]
\centering
\caption{System-level comparison on the CoW suite. All systems are given the
same ontology context.}
\label{tab:systems}
\begin{tabular}{lc}
\toprule
System & Pass@1 (\%) \\
\midrule
DeepSeek-Prover-V2 7B \cite{deepseekproverv2} & 8.9\% \\
Kimina-Prover 7B \cite{kiminaprover}          & 3.3\% \\
Goedel-Prover-V2 32B \cite{goedelproverv2}    & 1.1\% \\
Backbone LLM, no agents                       & 58.9\% \\
\textbf{LAMP}                          & \textbf{96.7\%} \\
\bottomrule
\end{tabular}
\end{table}

Table~\ref{tab:systems} reports the result. Despite identical ontology access,
the specialized provers solve almost nothing---8/90, 3/90, and 1/90
respectively---while the backbone LLM alone solves 58.9\% and LAMP reaches
96.7\%. A miniF2F sanity check on Goedel-Prover-V2 confirms it is a functioning
prover, producing valid in-distribution proofs (8B: $6/10$, 32B: $2/10$), so its
near-total failure on CoW is not a broken pipeline. The gap instead reflects
architecture: specialized provers are trained to emit tactics directly from a
fixed Mathlib-centered distribution and have no mechanism to consume retrieved ontology context, whereas a general-purpose LLM can read and apply it directly, and LAMP's agentic structure exploits it further still.
 
\subsection{Out-of-Domain Generalization}
\label{sec:eval-ood}

To evaluate LAMP's generalization beyond its target domain, we assess all three models on a balanced 32-problem subset of miniF2F \cite{minif2f}, drawn equally from four domains: algebra, competition mathematics, induction, and number theory (8 problems each). This balanced sampling ensures representative coverage across problem types rather than favoring any single category. A full evaluation across all 244 miniF2F problems was not feasible within our computational budget, as each theorem requires multiple Planner, Builder cycles with repeated MCP tool invocations and Lean REPL verification. This makes large-scale evaluation significantly expensive. The 32-problem balanced subset was designed to provide a principled and reproducible cross-section of LAMP's out-of-domain behaviour under these constraints.

\begin{table}[t]
\centering
\caption{Out-of-domain evaluation on a balanced 32-problem subset of
miniF2F (8 problems per domain). All results are Pass@1.}
\label{tab:minif2f}
\setlength{\tabcolsep}{6pt}
\small
\begin{tabular}{lcccc|c}
\toprule
Model & Algebra & Comp.\ Math & Induction & Num.\ Theory & \textbf{Overall} \\
\midrule
Backbone model          & 100   & 100  & 100  & 87.5 & \textbf{96.9} \\
DeepSeek V4 Pro    & 75.0  & 37.5 & 62.5 & 50.0 & \textbf{56.2} \\
Claude Sonnet 4.5  & 87.5  & 12.5 & 25.0 & 12.5 & \textbf{34.4} \\
\bottomrule
\end{tabular}
\end{table}

Table~\ref{tab:minif2f} reports the Pass@1 results across all three
models. Backbone model (Kimi K2.6) achieves 96.9\%, while DeepSeek V4 Pro achieves 56.2\% and
Claude Sonnet 4.5 achieves 34.4\%. The significant variance across 
models suggests that out-of-domain generalization in LAMP is strongly governed
by the  model's agentic tool-use capabilities rather than the framework
itself. Claude's drop from 88.9\% on CoW to 34.4\% on miniF2F further
indicates that its strong CoW performance is partly attributable to LAMP's
domain-specific ontology and semantic retrieval tools, which provide little
leverage on Mathlib-centered competition problems. We note that these Pass@1
results do not constitute a direct comparison with dedicated prover models
evaluated on the full miniF2F benchmark under pass@$k$ protocols; a
comprehensive comparison under unified evaluation conditions remains future
work.

\section{Discussion and Limitations}
\label{sec:discussion}
The above evaluation provides a quantitative overview of model performance; however, a deeper understanding requires examining where and why the models fail. Accordingly, we proceed with a detailed failure analysis of the backbone model to characterize its limitations and discuss future improvements.
\paragraph{Failure Analysis:}
An analysis of the failed proofs for Backbone model (Kimi K2.6) reveals distinct failure modes. The two failures in suffix\_of\_pow\_le and hasPeriod\_pow\_length stemmed from complex modular indexing lookups. Meanwhile, the failure in primitive\_minimal \_period\_is\_length occurred because the theorem statement within the dataset is mathematically false; the multi-agent Planner correctly identified this flaw and refused to generate a proof. 

\begin{table}[t]
\centering
\caption{Classification of LAMP's 38 failure instances on the CoW suite (Claude Sonnet 4.5 + DeepSeek v4 Pro combined).}
\label{tab:failure}
\small
\setlength{\tabcolsep}{4pt}
\begin{tabular}{@{}lrc@{}}
\toprule
Failure Category & Count & \% \\
\midrule
Incorrect proof strategy        & 29 & 76.3 \\
Invalid Lean syntax             &  7 & 18.4 \\
Ill-posed (false) theorem       &  2 &  5.3 \\
\bottomrule
\end{tabular}
\end{table}

We examined the 38 proof attempts that LAMP failed to complete using the Claude Sonnet 4.5 and DeepSeek v4 Pro backbones and classified the failures into the categories shown in Table~\ref{tab:failure}. The dominant category was incorrect proof strategy $(76.3\%)$, where the agent produced a plausible plan but could not close all proof obligations, typically leaving \texttt{sorry} placeholders after exhausting all re-planning attempts. Seven theorems, related to the Factor, Period, and Border modules, failed under both backbones, indicating that these proofs require multi-step modular-arithmetic reasoning over \texttt{List.get?} indices that exceeds current LLM capabilities regardless of backbone choice. The second category, invalid Lean syntax $(18.4\%)$, consisted entirely of DeepSeek failures on Easy and Medium theorems that both Kimi K2.6 and Claude Sonnet solved; the proof ideas were correct, but the generated code referenced non-existent identifiers or applied wrong projection syntax, suggesting lower Lean 4 surface-syntax fidelity for DeepSeek. Two failures $(5.3\%)$ stemmed from the same ill-posed theorem discussed above; neither Claude nor DeepSeek identified its falsity, unlike Kimi K2.6. Notably, no failures were attributable to verification timeouts or tool retrieval errors, confirming that LAMP's MCP-based infrastructure operates reliably and that the performance ceiling is determined by backbone reasoning quality rather than system-level bottlenecks. We now discuss the broader implications of these results, along with key limitations and  future directions. 

The central finding of our evaluation is that a general-purpose language model equipped with curated domain tooling outperforms specialized provers in a domain for which no training corpus exists, and that this advantage depends on LAMP's tool-grounded architecture and its Planner/Builder separation, not on the backbone model alone (Section~\ref{sec:eval-ablation}). We read this as evidence that, for unformalized domains, the binding constraint is not the model's reasoning capacity but its access to the domain's vocabulary and the relationships among its definitions. Specialized provers encode such knowledge implicitly, through training on a fixed Mathlib-centered corpus, and consequently degrade when that corpus does not cover the target domain \cite{taobench}. LAMP instead supplies the knowledge explicitly and at inference time,making it retargetable to new domains through library curation rather than model retraining. This gives a solution for formalizing further unformalized domains.
  
\paragraph{Limitations and Future Work:}
The CoW Evaluation Suite is self-constructed, and its difficulty labels reflect our own judgement rather than an external standard; pass rates on it are not directly comparable to established benchmarks.  As noted in Section~\ref{sec:eval-setup}, the suite is a foundation rather than an endpoint, and the same library-growth loop can extend it toward the field's hardest classical results, such as the Fine-Wilf theorem~\cite{fine1965uniqueness}.  More broadly, results routinely dismissed as ``obvious'' in informal mathematics (e.g.,\ the border-period duality, recorded only as a remark in~\cite{Simpson2025palindromic}) require explicit machine-checked proofs on which stronger statements depend. This gap between ``easy to see'' and ``accepted by the kernel'' is where much formalisation effort concentrates and where the domain ontology
contributes most.
LAMP's performance is also sensitive to the backbone model (Section~\ref{sec:eval-backbone}); the gap to open-source backbones is substantial, so LAMP is best understood as a method for
extracting domain-specific proving capability from a capable general-purpose model, not a model-agnostic guarantee.  This motivates a future hybrid design in which the Planner remains
general-purpose while the Builder is a prover-specialized model adapted to MCP tool calling, a split the Planner/Builder separation was designed to accommodate, since current specialized provers (e.g.\ Kimina-Prover~\cite{kiminaprover}, DeepSeek-Prover-V2~\cite{deepseekproverv2}) do not yet expose native tool-calling interfaces.
Finally, our comparison with specialized provers spans a single unformalized domain; whether explicit domain tooling and the library-growth loop transfer to other such domains remains open, and is the natural next test of this methodology.

\section{Conclusion}
\label{sec:conclusion}

We presented LAMP, a multi-agent framework that synthesizes kernel-verified
Lean 4 proofs by supplying curated domain knowledge to a general-purpose
language model at inference time, together with a Lean 4 formalization of CoW, a domain previously absent from Mathlib, on which we evaluated it. Across a suite of 90 CoW theorems spanning eight modules and three difficulty levels, LAMP synthesized verified proofs for 96.7\% of theorems, and an ablation showed that this performance depends on LAMP's tool-grounded architecture rather than the backbone model alone. These results indicate that for domains lacking a formalized corpus, the decisive factor is explicit, structured access to domain knowledge, which a tool-augmented agent can exploit without retraining. The suite's 87 verified proofs are themselves candidates for the library. Folding them back would grow it from 93 to as many as 180 declarations, the kind of compounding foundation LAMP is designed to build as it solves. We release the CoW library and evaluation suite to support this and the further work.

\clearpage
\bibliographystyle{splncs04}
\bibliography{ref} 

\appendix
\section*{Appendix}
\addcontentsline{toc}{section}{Appendix}

\section{The CoW Library: Additional Details}
\label{app:cow}

\subsection*{Design rationale}
Words are represented as \texttt{List $\alpha$} rather than a custom inductive
type, so that all of Lean's existing list lemmas and automation are available
for free. \emph{Factor} is defined contiguously (\texttt{IsFactor u w := $\exists$
x y, w = x ++ u ++ y}) rather than via an indexed substring predicate, which
keeps proofs by case analysis on $x$ and $y$ direct rather than requiring index
arithmetic. All declarations reported in Table~\ref{tab:cow-library} are
sorry-free.

\subsection*{Notation}
\begin{center}
\begin{tabular}{lll}
\toprule
Notation & Meaning & Definition \\
\midrule
$u \cdot v$      & Concatenation   & \texttt{u ++ v} \\
$u \le_p w$      & Prefix          & $\exists s,\ w = u ++ s$ \\
$u \le_s w$      & Suffix          & $\exists p,\ w = p ++ u$ \\
$u \le_f w$      & Factor          & $\exists x\, y,\ w = x ++ u ++ y$ \\
$u <_p w$        & Proper Prefix   & $u \le_p w \land u \ne w$ \\
$u <_s w$        & Proper Suffix   & $u \le_s w \land u \ne w$ \\
$u \sim v$       & Conjugacy       & $\exists x\, y,\ u = x \cdot y \land v = y \cdot x$ \\
\bottomrule
\end{tabular}
\end{center}

\subsection*{Module summaries}
\emph{Word} contributes length properties (e.g.\ \texttt{Word.length\_concat}),
concatenation algebra (e.g.\ \texttt{Word.concat\_assoc}), and power laws (e.g.\ \texttt{Word.pow\_add}). \emph{Factor}
establishes reflexivity, transitivity, and length monotonicity for the
prefix, suffix, and factor relations, together with closure lemmas relating
the three (e.g.\ \texttt{IsFactor.of\_prefix}). \emph{ProperPrefix} and
\emph{ProperSuffix} each add the corresponding strict-inequality lemmas and
the empty-word boundary case. \emph{Border} adds \texttt{IsUnbordered} and
\texttt{IsLongestBorder} alongside \texttt{IsBorder.isFactor}
(Section~\ref{sec:cow}) and \texttt{not\_isBorder\_self}. \emph{Conjugacy}
establishes reflexivity, symmetry, and length preservation under cyclic
rotation. \emph{Period} includes the trivial full-length period
(\texttt{hasPeriod\_length}) and an equivalence characterization of
primitivity (\texttt{isPrimitive\_iff}). \emph{Morphism} covers application
over concatenation and powers, composition, classification relationships
between coding, uniform, and non-erasing morphisms, and preservation of
factor, prefix, and suffix relations under morphism application.

\section{Evaluation suite: Additional details}
\label{app:eval}
Table \ref{tab:cow_evaluation} lists every theorem in the suite together with its module, difficulty label, and per-backbone pass/fail outcomes.
\begin{longtable}{rlllccc}
\caption{Evaluation results on the CoW suite, per theorem.}
\label{tab:cow_evaluation} \\
\toprule
\textbf{\#} & \textbf{Theorem} & \textbf{Module} & \textbf{Type} & \textbf{Kimi} & \textbf{Claude} & \textbf{DeepSeek} \\
\midrule
\endfirsthead
\multicolumn{7}{l}{\small\textit{Table~\ref{tab:cow_evaluation} continued}} \\
\toprule
\textbf{\#} & \textbf{Theorem} & \textbf{Module} & \textbf{Type} & \textbf{Kimi} & \textbf{Claude} & \textbf{DeepSeek} \\
\midrule
\endhead
\midrule
\multicolumn{7}{r}{\small\textit{continued on next page}} \\
\endfoot
\bottomrule
\endlastfoot
1 & \texttt{word\_length\_eq\_zero\_iff} & Word & Easy & $\checkmark$ & $\checkmark$ & $\checkmark$ \\
2 & \texttt{word\_length\_pos\_of\_ne\_nil} & Word & Easy & $\checkmark$ & $\checkmark$ & $\checkmark$ \\
3 & \texttt{word\_concat\_right\_cancel} & Word & Medium & $\checkmark$ & $\checkmark$ & $\checkmark$ \\
4 & \texttt{word\_concat\_left\_cancel} & Word & Medium & $\checkmark$ & $\checkmark$ & $\checkmark$ \\
5 & \texttt{word\_pow\_two\_length} & Word & Medium & $\checkmark$ & $\checkmark$ & $\checkmark$ \\
6 & \texttt{word\_pow\_succ\_left} & Word & Medium & $\checkmark$ & $\checkmark$ & $\checkmark$ \\
7 & \texttt{pow\_ne\_empty\_of\_ne\_empty} & Word & Medium & $\checkmark$ & $\checkmark$ & $\checkmark$ \\
8 & \texttt{concat\_eq\_nil\_iff} & Word & Medium & $\checkmark$ & $\checkmark$ & $\checkmark$ \\
9 & \texttt{pow\_eq\_nil\_iff} & Word & Medium & $\checkmark$ & $\times$ & $\checkmark$ \\
10 & \texttt{pow\_comm\_concat} & Word & Medium & $\checkmark$ & $\checkmark$ & $\checkmark$ \\
11 & \texttt{pow\_injective\_of\_ne\_empty} & Word & Medium & $\checkmark$ & $\checkmark$ & $\checkmark$ \\
12 & \texttt{pow\_add\_length\_eq} & Word & Medium & $\checkmark$ & $\checkmark$ & $\checkmark$ \\
13 & \texttt{prefix\_of\_concat} & Factor & Easy & $\checkmark$ & $\checkmark$ & $\checkmark$ \\
14 & \texttt{suffix\_of\_concat} & Factor & Easy & $\checkmark$ & $\checkmark$ & $\checkmark$ \\
15 & \texttt{prefix\_of\_factor\_is\_factor} & Factor & Medium & $\checkmark$ & $\checkmark$ & $\checkmark$ \\
16 & \texttt{prefix\_antisymm} & Factor & Medium & $\checkmark$ & $\checkmark$ & $\times$ \\
17 & \texttt{suffix\_antisymm} & Factor & Medium & $\checkmark$ & $\times$ & $\times$ \\
18 & \texttt{suffix\_of\_factor\_is\_factor} & Factor & Medium & $\checkmark$ & $\checkmark$ & $\checkmark$ \\
19 & \texttt{prefix\_concat\_right} & Factor & Medium & $\checkmark$ & $\checkmark$ & $\checkmark$ \\
20 & \texttt{suffix\_concat\_left} & Factor & Medium & $\checkmark$ & $\checkmark$ & $\checkmark$ \\
21 & \texttt{factor\_of\_concat\_left} & Factor & Medium & $\checkmark$ & $\checkmark$ & $\checkmark$ \\
22 & \texttt{factor\_of\_concat\_right} & Factor & Medium & $\checkmark$ & $\checkmark$ & $\checkmark$ \\
23 & \texttt{prefix\_of\_pow\_succ} & Factor & Medium & $\checkmark$ & $\checkmark$ & $\checkmark$ \\
24 & \texttt{prefix\_of\_pow\_le} & Factor & Medium & $\checkmark$ & $\checkmark$ & $\checkmark$ \\
25 & \texttt{suffix\_of\_pow\_le} & Factor & Medium & $\times$ & $\times$ & $\checkmark$ \\
26 & \texttt{prefix\_eq\_of\_length\_eq} & Factor & Hard & $\checkmark$ & $\checkmark$ & $\checkmark$ \\
27 & \texttt{prefix\_suffix\_concat\_partition} & Factor & Hard & $\checkmark$ & $\times$ & $\times$ \\
28 & \texttt{proper\_prefix\_irrefl} & ProperPrefix & Easy & $\checkmark$ & $\checkmark$ & $\checkmark$ \\
29 & \texttt{proper\_prefix\_ne} & ProperPrefix & Easy & $\checkmark$ & $\checkmark$ & $\checkmark$ \\
30 & \texttt{prefix\_lt\_of\_length\_lt} & ProperPrefix & Medium & $\checkmark$ & $\checkmark$ & $\checkmark$ \\
31 & \texttt{proper\_prefix\_trans} & ProperPrefix & Medium & $\checkmark$ & $\checkmark$ & $\times$ \\
32 & \texttt{proper\_prefix\_concat\_ne\_nil} & ProperPrefix & Medium & $\checkmark$ & $\checkmark$ & $\checkmark$ \\
33 & \texttt{proper\_prefix\_of\_prefix\_of\_proper\_prefix} & ProperPrefix & Medium & $\checkmark$ & $\checkmark$ & $\checkmark$ \\
34 & \texttt{proper\_prefix\_of\_proper\_prefix\_of\_prefix} & ProperPrefix & Medium & $\checkmark$ & $\checkmark$ & $\checkmark$ \\
35 & \texttt{proper\_prefix\_witness} & ProperPrefix & Medium & $\checkmark$ & $\checkmark$ & $\times$ \\
36 & \texttt{proper\_prefix\_witness\_is\_suffix} & ProperPrefix & Medium & $\checkmark$ & $\checkmark$ & $\checkmark$ \\
37 & \texttt{proper\_prefix\_double\_lt} & ProperPrefix & Medium & $\checkmark$ & $\checkmark$ & $\checkmark$ \\
38 & \texttt{proper\_suffix\_irrefl} & ProperSuffix & Easy & $\checkmark$ & $\checkmark$ & $\checkmark$ \\
39 & \texttt{proper\_suffix\_ne} & ProperSuffix & Easy & $\checkmark$ & $\checkmark$ & $\checkmark$ \\
40 & \texttt{suffix\_lt\_of\_length\_lt} & ProperSuffix & Medium & $\checkmark$ & $\checkmark$ & $\checkmark$ \\
41 & \texttt{proper\_suffix\_trans} & ProperSuffix & Medium & $\checkmark$ & $\checkmark$ & $\times$ \\
42 & \texttt{proper\_suffix\_concat\_ne\_nil} & ProperSuffix & Medium & $\checkmark$ & $\checkmark$ & $\checkmark$ \\
43 & \texttt{proper\_suffix\_of\_suffix\_of\_proper\_suffix} & ProperSuffix & Medium & $\checkmark$ & $\checkmark$ & $\checkmark$ \\
44 & \texttt{proper\_suffix\_of\_proper\_suffix\_of\_suffix} & ProperSuffix & Medium & $\checkmark$ & $\checkmark$ & $\times$ \\
45 & \texttt{proper\_suffix\_witness} & ProperSuffix & Medium & $\checkmark$ & $\checkmark$ & $\checkmark$ \\
46 & \texttt{proper\_suffix\_witness\_is\_prefix} & ProperSuffix & Medium & $\checkmark$ & $\checkmark$ & $\checkmark$ \\
47 & \texttt{proper\_suffix\_double\_lt} & ProperSuffix & Medium & $\checkmark$ & $\checkmark$ & $\checkmark$ \\
48 & \texttt{border\_is\_proper\_prefix} & Border & Easy & $\checkmark$ & $\checkmark$ & $\checkmark$ \\
49 & \texttt{border\_is\_proper\_suffix} & Border & Easy & $\checkmark$ & $\checkmark$ & $\checkmark$ \\
50 & \texttt{border\_ne\_word} & Border & Medium & $\checkmark$ & $\checkmark$ & $\checkmark$ \\
51 & \texttt{border\_empty\_iff} & Border & Medium & $\checkmark$ & $\checkmark$ & $\checkmark$ \\
52 & \texttt{longest\_border\_length\_unique} & Border & Medium & $\checkmark$ & $\checkmark$ & $\checkmark$ \\
53 & \texttt{not\_unbordered\_iff\_exists\_nonempty\_border} & Border & Medium & $\checkmark$ & $\checkmark$ & $\checkmark$ \\
54 & \texttt{not\_border\_of\_length\_ge} & Border & Medium & $\checkmark$ & $\checkmark$ & $\checkmark$ \\
55 & \texttt{border\_of\_prefix\_suffix} & Border & Medium & $\checkmark$ & $\checkmark$ & $\times$ \\
56 & \texttt{border\_nonempty\_implies\_length\_ge\_two} & Border & Medium & $\checkmark$ & $\checkmark$ & $\checkmark$ \\
57 & \texttt{border\_imp\_not\_unbordered} & Border & Medium & $\checkmark$ & $\checkmark$ & $\checkmark$ \\
58 & \texttt{border\_of\_border\_is\_border} & Border & Hard & $\checkmark$ & $\checkmark$ & $\checkmark$ \\
59 & \texttt{border\_period\_duality} & Border & Hard & $\checkmark$ & $\times$ & $\times$ \\
60 & \texttt{conjugate\_concat} & Conjugacy & Easy & $\checkmark$ & $\checkmark$ & $\checkmark$ \\
61 & \texttt{conjugate\_of\_eq} & Conjugacy & Easy & $\checkmark$ & $\checkmark$ & $\checkmark$ \\
62 & \texttt{conjugate\_symm\_iff} & Conjugacy & Medium & $\checkmark$ & $\checkmark$ & $\checkmark$ \\
63 & \texttt{conjugate\_empty\_iff} & Conjugacy & Medium & $\checkmark$ & $\checkmark$ & $\checkmark$ \\
64 & \texttt{conjugate\_concat\_rotate} & Conjugacy & Medium & $\checkmark$ & $\checkmark$ & $\checkmark$ \\
65 & \texttt{conjugate\_preserves\_ne\_empty} & Conjugacy & Medium & $\checkmark$ & $\checkmark$ & $\checkmark$ \\
66 & \texttt{conjugate\_pow\_rotate} & Conjugacy & Medium & $\checkmark$ & $\checkmark$ & $\checkmark$ \\
67 & \texttt{conjugate\_trans} & Conjugacy & Hard & $\checkmark$ & $\checkmark$ & $\times$ \\
68 & \texttt{conjugate\_pow\_of\_conjugate} & Conjugacy & Hard & $\checkmark$ & $\times$ & $\checkmark$ \\
69 & \texttt{primitive\_length\_pos} & Period & Easy & $\checkmark$ & $\checkmark$ & $\times$ \\
70 & \texttt{hasPeriod\_of\_short} & Period & Easy & $\checkmark$ & $\checkmark$ & $\times$ \\
71 & \texttt{isPrimitive\_of\_length\_one} & Period & Medium & $\checkmark$ & $\checkmark$ & $\times$ \\
72 & \texttt{minimalPeriod\_unique} & Period & Medium & $\checkmark$ & $\checkmark$ & $\times$ \\
73 & \texttt{pow\_two\_not\_primitive} & Period & Medium & $\checkmark$ & $\checkmark$ & $\checkmark$ \\
74 & \texttt{primitive\_minimal\_period\_is\_length} & Period & Hard & $\times$ & $\times$ & $\times$ \\
75 & \texttt{hasPeriod\_concat\_self} & Period & Hard & $\checkmark$ & $\times$ & $\times$ \\
76 & \texttt{hasPeriod\_mul\_of\_hasPeriod} & Period & Hard & $\checkmark$ & $\checkmark$ & $\times$ \\
77 & \texttt{period\_of\_prefix} & Period & Hard & $\checkmark$ & $\times$ & $\times$ \\
78 & \texttt{hasPeriod\_pow\_length} & Period & Hard & $\times$ & $\times$ & $\times$ \\
79 & \texttt{morphism\_cons} & Morphism & Easy & $\checkmark$ & $\checkmark$ & $\times$ \\
80 & \texttt{coding\_preserves\_length} & Morphism & Medium & $\checkmark$ & $\checkmark$ & $\times$ \\
81 & \texttt{identity\_morphism} & Morphism & Medium & $\checkmark$ & $\checkmark$ & $\times$ \\
82 & \texttt{nonErasing\_preserves\_ne\_empty} & Morphism & Medium & $\checkmark$ & $\checkmark$ & $\times$ \\
83 & \texttt{uniform\_pow\_length} & Morphism & Medium & $\checkmark$ & $\checkmark$ & $\times$ \\
84 & \texttt{morphism\_length\_cons} & Morphism & Medium & $\checkmark$ & $\checkmark$ & $\checkmark$ \\
85 & \texttt{comp\_nonErasing} & Morphism & Hard & $\checkmark$ & $\checkmark$ & $\times$ \\
86 & \texttt{coding\_comp\_coding} & Morphism & Hard & $\checkmark$ & $\checkmark$ & $\checkmark$ \\
87 & \texttt{uniform\_comp} & Morphism & Hard & $\checkmark$ & $\checkmark$ & $\times$ \\
88 & \texttt{morphism\_preserves\_proper\_prefix\_ne} & Morphism & Hard & $\checkmark$ & $\checkmark$ & $\times$ \\
89 & \texttt{morphism\_preserves\_proper\_suffix\_ne} & Morphism & Hard & $\checkmark$ & $\checkmark$ & $\checkmark$ \\
90 & \texttt{border\_preserved\_by\_coding} & Morphism & Hard & $\checkmark$ & $\checkmark$ & $\times$ \\
\midrule
\multicolumn{4}{l}{\textbf{Pass@1 (Total: 90)}} & \textbf{96.7\%} & \textbf{88.9\%} & \textbf{68.9\%} \\
\end{longtable}

\begin{table}[]
\centering
\caption{LAMP pass@1 on the CoW Evaluation Suite, by module and backbone.}
\label{tab:main-module}
\begin{tabular}{lcccc}
\toprule
Module & \#\,Theorems & Kimi & Claude & DeepSeek \\
\midrule
Word         & 12 & 100.0\% &  91.7\% & 100.0\% \\
Factor       & 15 &  93.3\% &  80.0\% &  80.0\% \\
ProperPrefix & 10 & 100.0\% & 100.0\% &  80.0\% \\
ProperSuffix & 10 & 100.0\% & 100.0\% &  80.0\% \\
Border       & 12 & 100.0\% &  91.7\% &  83.3\% \\
Conjugacy    &  9 & 100.0\% &  88.9\% &  88.9\% \\
Period       & 10 &  80.0\% &  60.0\% &  10.0\% \\
Morphism     & 12 & 100.0\% & 100.0\% &  25.0\% \\
\midrule
\textbf{Overall} & \textbf{90} & \textbf{96.7\%} & \textbf{88.9\%} & \textbf{68.9\%} \\
\bottomrule
\end{tabular}
\end{table}

\section{Representative result: border--period duality}
\label{app:results}

To illustrate LAMP's input/output behavior concretely, we show the unsolved
input and verified output for \texttt{border\_period\_duality}
(Table~\ref{tab:cow_evaluation}, \#59), the classical result that a border of
length $|b|$ induces a period of length $|w| - |b|$ \cite{lothaire2002algebraic}. The Orchestrator's edit
localization (Section~\ref{sec:lamp-components}) means only the proof body
changes between input and output; the theorem signature is never touched.

The unsolved input provided to the proposed system is as follows:
\begin{lstlisting}
theorem border_period_duality {b w : Word α} (h : IsBorder b w) :
    HasPeriod w (w.length - b.length) := by
  sorry
\end{lstlisting}

LAMP patches only the region after \texttt{:= by}, producing the following
kernel-verified proof, which chains a prefix decomposition and a suffix
decomposition of $w$ through the \emph{Border}, \emph{Factor}, and
\emph{Period} modules:

\begin{lstlisting}
theorem border_period_duality {b w : Word α} (h : IsBorder b w) :
    HasPeriod w (w.length - b.length) := by
  have h1 : b <ₚ w := h.1
  have h2 : b <ₛ w := h.2
  obtain ⟨x, hw1, hxne⟩ := (IsProperPrefix.def b w).mp h1
  obtain ⟨y, hw2, hyne⟩ := (IsProperSuffix.iff b w).mp h2
  have hlen1 : w.length = b.length + x.length := by
    rw [hw1]; simp [List.length_append]
  have hlen2 : w.length = y.length + b.length := by
    rw [hw2]; simp [List.length_append]
  have hp_eq_xlen : w.length - b.length = x.length := by
    rw [hlen1]; omega
  have hp_eq_ylen : w.length - b.length = y.length := by
    rw [hlen2]; omega
  constructor
  · rw [hp_eq_xlen]
    apply List.length_pos_iff_ne_nil.mpr
    exact hxne
  · intro i hi
    have hib : i < b.length := by omega
    have hw_i : w.get? i = b.get? i := by
      rw [hw1]; simp [List.getElem?_append, hib]
    have hw_ip : w.get? (i + (w.length - b.length)) = b.get? i := by
      rw [hp_eq_ylen, hw2]
      have hiy : i + y.length ≥ y.length := by omega
      have hget : (y ++ b).get? (i + y.length)
          = b.get? (i + y.length - y.length) := by
        simp [List.getElem?_append, hiy]
      rw [hget]
      have hid : i + y.length - y.length = i := by omega
      rw [hid]
    exact hw_i.trans hw_ip.symm
\end{lstlisting}

Note that the border-period duality is universally treated as a trivial remark in the literature (e.g.,\ Remark~1.1 in~\cite{Simpson2025palindromic}; ``well known'' in~\cite{lothaire2002algebraic}), yet a machine-checked proof requires an explicit witness linking the prefix and suffix overlap to the period offset.  The proposed
formalisation above makes this step precise.

\end{document}